\documentclass[
aps,%
12pt,%
final,%
notitlepage,%
oneside,%
onecolumn,%
nobibnotes,%
nofootinbib,
superscriptaddress,%
noshowpacs,%
centertags]%
{revtex4}

\begin{document}
\selectlanguage{english} 
\title{The hydrogen and helium lines\\ of the symbiotic binary Z~And during its brightening\\ at the end of 2002 
\footnote{Based on observations collected at the National Astronomical Observatory Rozhen, Bulgaria}}
\author{\firstname{N. A.} \surname{Tomov}}
\email[]{tomov@astro.bas.bg}
\affiliation{Institute of Astronomy, Bulgarian Academy of Science, National Astronomical Observatory Rozhen,\\ POBox 136, 4700 Smolyan, Bulgaria\\}
\author{\firstname{M. T.} \surname{Tomova}}
\email[]{mtomova@astro.bas.bg}
\affiliation{Institute of Astronomy, Bulgarian Academy of Science, National Astronomical Observatory Rozhen,\\ POBox 136, 4700 Smolyan, Bulgaria\\}
\author{\firstname{D. V.} \surname{Bisikalo}}
\email[]{bisikalo@inasan.rssi.ru}
\affiliation{Institute of Astronomy of the Russian Academy of Science, 48 Pyatnitskaya Str., 119017 Moscow, Russia}

\begin{abstract}
High resolution observations in the region of the lines H$\alpha$,
\mbox{He\,{\sc ii}} $\lambda$\,4686 and H$\gamma$ of the spectrum of
the symbiotic binary Z~And were performed during its
small-amplitude brightening at the end of 2002. The profiles of the
hydrogen lines were double-peaked. These profiles give a reason to
suppose that the lines can be emitted mainly by an optically thin
accretion disc. The H$\alpha$ line is strongly contaminated by the
emission of the envelope, therefore for consideration of accretion disc
properties we use the H$\gamma$ line. The H$\alpha$ line had broad
wings which are supposed to be determined mostly from radiation damping 
but high velocity stellar wind
from the compact object in the system can also contribute to their appearance. 
The H$\gamma$ line had a broad emission component which is assumed to be emitted mainly from the inner part of the accretion disc.
The line \mbox{He\,{\sc ii}} $\lambda$\,4686 had a broad emission component too, but it is supposed to appear in a region of a high velocity stellar wind. The outer radius of the
accretion disc can be calculated from the shift between the peaks. Assuming,
that the orbit inclination can ranges from 47$^\circ$ to 76$^\circ$
we estimate the outer radius as 20 -- 50 R$_{\odot}$. The behaviour
of the observed lines can be considered in the framework of the model
proposed for interpretation of the line spectrum during the major
2000 -- 2002 brightening of this binary. 
\end{abstract}

\maketitle
\section{Introduction}

Symbiotic stars are interpreted as interacting binaries consisting
of a cool giant of luminosity type III - II and a compact component
accreting matter from the atmosphere of the giant. As a result of
accretion the compact object undergoes multiple eruptions
accompanied by intensive loss of mass. Symbiotic stars provide a
good opportunity to study the processes of accretion and loss of
mass which are spectroscopically observed and form multicomponent
profiles of their spectral lines. The system Z~And consists of a
normal cool giant of spectral type M4.5 \citep{MS}, hot compact
component with temperature higher than 10$^5$ K \citep{FC88,Sok06}
and extended circum-binary nebula formed by the winds of two
components. The orbital period of this binary is 758.8$^{\rm d}$,
derived from both photometric \citep{FL} and radial velocity
\citep{MK96} data.

Z~And has undergone several active phases (after 1915, 1939, 1960,
1984 and 2000) consisting of repeated optical brightenings with
amplitudes up to 2--3 mag which are characterized by intensive loss
of mass \citep*{SS,Boyarchuk,FC95,TTT03,Sk06,Sok06,TTB08,Sk09}. The
last active phase of Z~And began at the end of August 2000
\citep{Sk00} and includes five optical brightenings. The second of
them developed after August 2002, when the light began to rise after
a deep minimum. The light reached its maximum in November and after
that gradually decreased reaching its quiescent value in the middle
of 2003. \citet{Sk06} analyzed the wings of the H$\alpha$ line and
concluded on this base that there was intensive loss of mass by the
compact object during that brightening.

In our previous work \citep*{TTT04} we analyzed the continuum
emission of Z~And during its 2002 brightening using multicolour
photometric data. We concluded that the increase of the fluxes in
the region $UBVRJHKL$ was mainly due to the nebular emission. The
energy distribution of the secondary component changed little (but
not negligible), it remained a hot compact object as in the
quiescent state of the system. To explain the line spectrum of the
system during the major 2000 -- 2002 brightening a gas dynamic model
was proposed where the high-velocity stellar wind of the compact
object having appeared at that time collides with the accretion disc
and an optically thick disc-like shell forms \citep{TTB08}. The two
velocity mass outflow as well as the behavior of the line
\mbox{He\,{\sc ii}} $\lambda$\,4686 during that brightening were
interpreted in the framework of this model. An emission detail of
the quiescent profile of the line H$\gamma$ was interpreted as a
component emitted by the accretion disc in the system.

During the growth of the light in the period September -- November
2002 the emission line spectrum of the system had various features.
The H$\alpha$ and H$\gamma$ lines had double-peaked profile. The
H$\alpha$ line had broad wings extended to not less than $\pm
2000$~km\,s$^{-1}$ from its center. The lines H$\gamma$ and
\mbox{He\,{\sc ii}} $\lambda$\,4686 had a broad emission component
with a low intensity which indicated velocities up to more than 1000
km\,s$^{-1}$. This paper is devoted to an analysis of the line
profiles of Z~And during its brightening at the end of 2002. Our aim
is to evaluate the parameters of the accretion disc, to conclude
about the possible nature of the nebular formations surrounding the
hot compact component at that time, and to examine the possibility
the emission line spectrum to be explained in the framework of the
same gas dynamic model.

\section{Observations and reduction}

High resolution data in the regions of the lines H$\alpha$,
\mbox{He\,{\sc ii}} $\lambda$\,4686 and H$\gamma$ of the spectrum of
the Z~And system were obtained in 2002 with a Photometrics CCD
camera mounted on the Coude spectrograph of the 2m RCC telescope of
the National Astronomical Observatory Rozhen (Table~1, Fig.~1). The
spectral resolution was 0.2 \AA\,px$^{-1}$ on all occasions.

\begin{table}

{\bf Table 1.} List of the observations and the H$\alpha$ flux in units $10^{-12}$ erg\,cm$^{-2}$\,s$^{-1}$.

  \label{list+Ha}
\center \begin{tabular}{|c|c|c|c|}
  \hline
    Date & JD$-$ & Orb. & H$\alpha$ \\
         & 2\,452\,000 & phase & flux \\

    \hline
 Sept. 25 & 543.35 & 0.017 & 173.494 \\
 Oct. 20 & 568.37 & 0.050 & 255.255 \\
 Nov. 12 & 591.20 & 0.080 & 292.098 \\
  \hline
 \end{tabular}
\end{table}

\begin{figure}
\begin{minipage}{8.4cm}
\resizebox{\hsize}{!}{\includegraphics[]{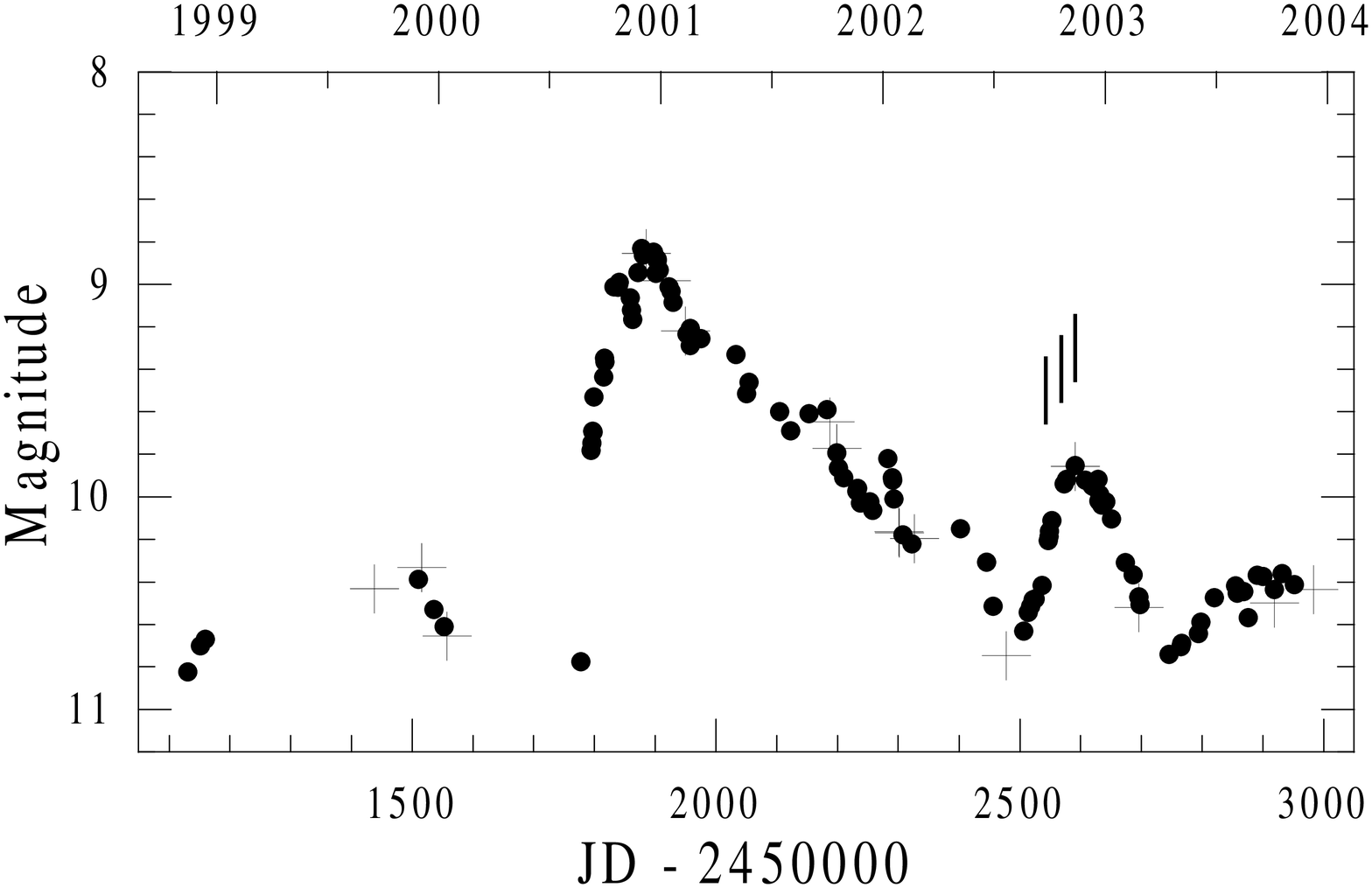}}
\end{minipage}

{\bf Figure 1.} The $V$ light curve of Z~And where the first two brightenings of its last active phase are seen.
The dots indicate the data of \citet{Sk02,Sk04} and the crosses -- our data. The vertical lines indicate the epochs
of the spectral observations.


\label{light curve}
\end{figure}

The initial data reduction and calculation of the line fluxes was
described in our previous work \citep{TTB08}. The continuum flux at
the wavelength position of the H$\alpha$ line was calculated using
linear extrapolation of the $V$ and $R$ photometric fluxes taken on
the same night or close nights. To calculate the fluxes of the
spectra from September and October we used photometric data of
\citet{Sk04}. The fluxes of the spectra obtained in November were
calculated on the basis of the photometric data for November 12,
2002 from our work \citet{TTT04}.

The $BV$ fluxes were corrected for the strong emission lines of Z~And in the same way as the quiescent data in the
work \citet{TTT03} because of the fact that the heights of these lines during the active phase were practically the
same as those in the quiescence.

All the fluxes were corrected for the interstellar reddening of
$E(B-V) = 0.30$ using the extinction law of \citet{Seaton}.

As in our previous papers on Z~And \citep{TTT03,TTT04,TTB08} we used
an ephemeris $\rm {Min(vis)=JD}~2\,442\,666^{\rm d}+758.8^{\rm d}
\times E$, where the orbital period is based on both photometric and
spectral data and the epoch of the orbital photometric minimum
coincides with that of the spectral conjunction
\citep{FL,MK96,Fekel2}.

    \section[]{Analysis of the emission line spectrum}

        \subsection[]{The Balmer lines}
        \subsubsection[]{The H$\alpha$ line}

In the quiescent state of the system the line H$\alpha$ was of the
nebular type, but had, in addition, broad wings extended to not less
than $\pm 2000$~km\,s$^{-1}$ from its center. It exceeded the local
continuum by a factor of more than 100 at orbital phases close to
0.5 \citep{TTB08}. The width (FWHM) of the line was about 100--120
km\,s$^{-1}$ and only occasionally went beyond this value. The width
remained the same during the 2000 -- 2002 optical brightening too
\citep{TTB08}. The view that the broad wings of the line in the
quiescent state of the system are due to Raman scattering of
Ly$\beta$ photons by atomic hydrogen is widely accepted
\citep{Lee,ATP,TTB08}. \citet{Lee}, though, considered other
theoretical possibilities for their appearance too. He noted that Raman scattering and radiation damping depend on the wavelength in the same way, and the latter also cannot be excluded.

\begin{figure*}
    \includegraphics[height=.45\textheight]{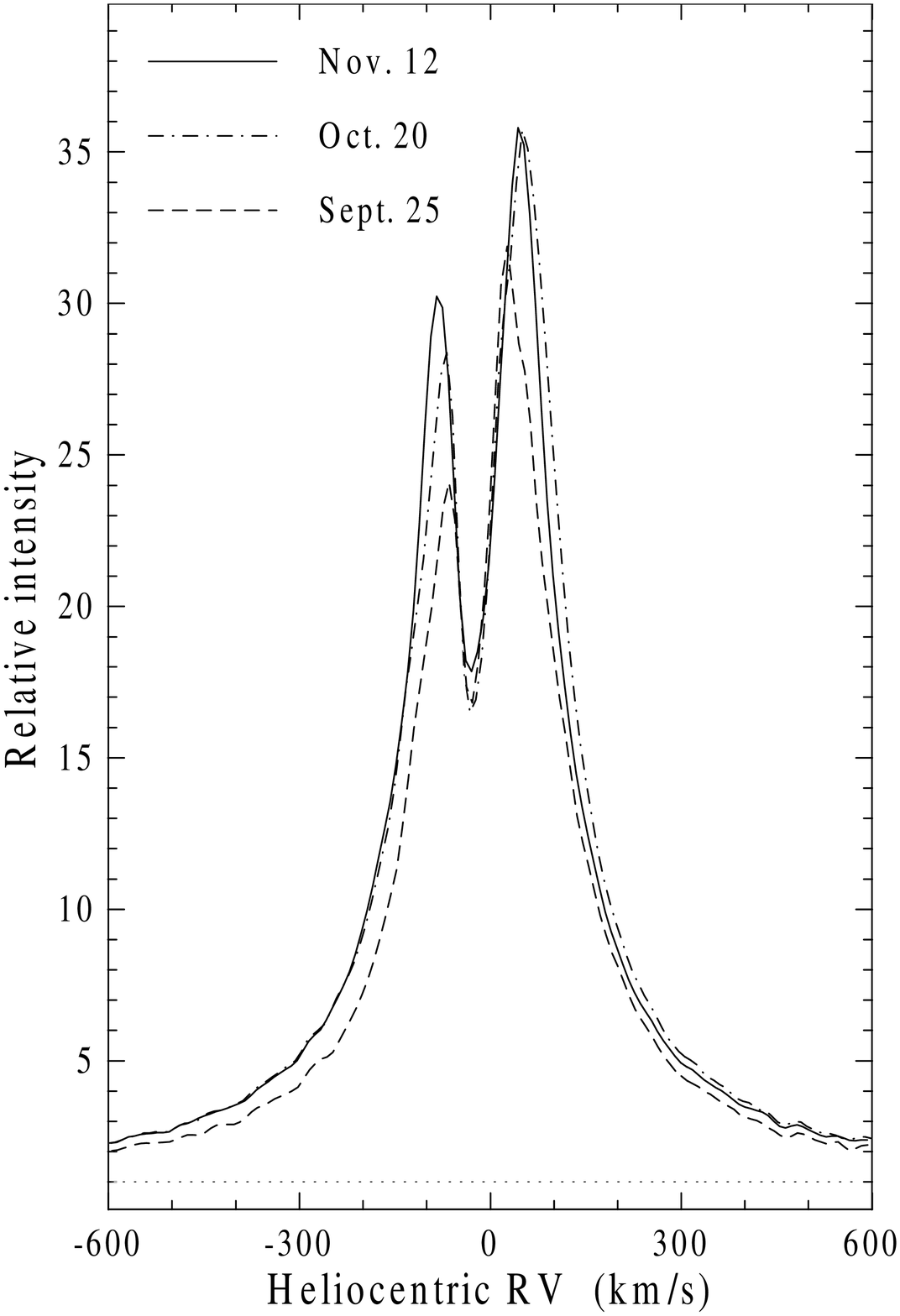}
    \includegraphics[height=.45\textheight]{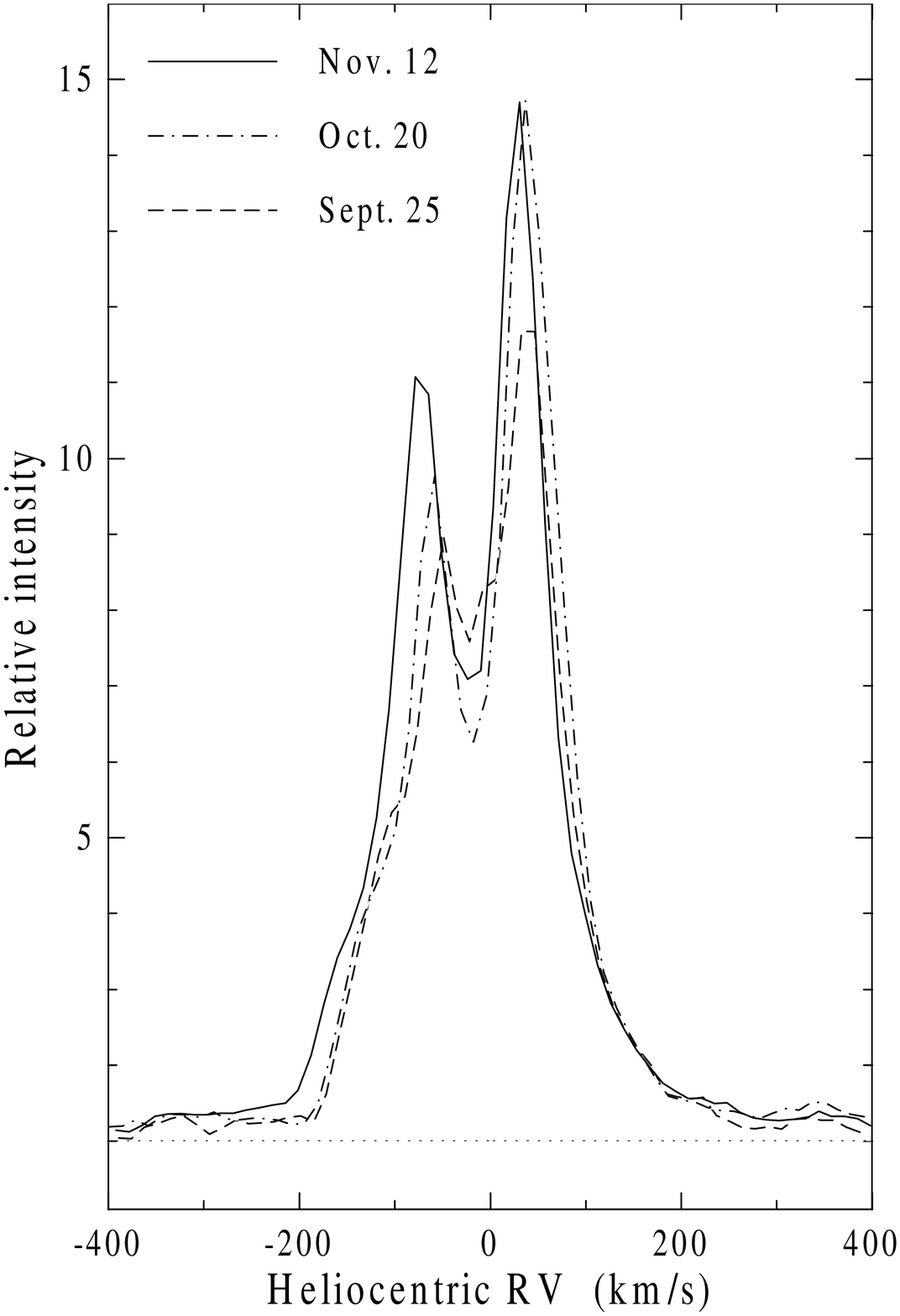}

{\bf Figure 2.} The profiles of the H$\alpha$ (left panel) and H$\gamma$ (right panel) lines. The level of the
local continuum is marked with a doted line.

\label{Ha&Hg prof}
\end{figure*}

During the growth of the light at the end of 2002 the line was double (Fig.~2), it exceeded the local
continuum by a factor of about 30 and its width increased by a factor of two. The dip between the two peaks had the
same radial velocity at all times of observation.

Let us compare the H$\alpha$ flux in December 2000 at the time of
the light maximum of the first brightening \citep{TTB08} and in the
light maximum in November 2002 (Table~1). The two maxima are at
close orbital phases -- 0.15 and 0.08. During the first brightening
the emission measure of the circum-binary nebula was greater by a
factor of two compared with the second brightening
\citep{TTT03,TTT04}, which means that the quantity of the emitting
gas has been greater by a factor of about two. We suppose close
physical conditions in the nebula during these brightenings because,
in the first place, the same mean electron temperature of 20\,000
K\citep{TTT03,TTT04} was obtained. Moreover, the mean electron
density during the first brightening will be higher only by a factor
of $\sqrt{2}$ if the nebula has the same volume.

If all H$\alpha$ photons left the nebula, the H$\alpha$ flux during
the first brightening would be greater by a factor at least of two.
The H$\alpha$ flux however, was approximately the same -- about (270
-- 290) $\times$ $10^{-12}$ erg\,cm$^{-2}$\,s$^{-1}$ in the two
light maxima. This means that during the first brightening the
optical depth in this line has been greater. When all photons leave
the emission region of one nebular line its profile is very close to
Gaussian. The more photons are absorbed, the more the line profile
will differ from the Gaussian function and can become a
double-peaked one. If we suppose that the double-peaked profile
during the second brightening is due only to the optical depth it
must be greater than the depth during the first one when the profile
was not double-peaked. The contradiction can be removed if we
suppose that the double-peaked profile is mainly due to kinematics
of the gas -- one possibility is an emission of an accretion disc.
The same conclusion is obtained when treating the line H$\gamma$.

If we suppose that the H$\alpha$ line is emitted mainly by an
accretion disc we must expect that the intensity of this line
greatly increases during the brightening. The H$\alpha$ emission in
this case is a sum of the emission of the circum-binary nebula and
the disc. The emission of the nebula can increase or at least will
not decrease during the brightening. At the same time disc's
emission must predominate over that of the nebula to determine the
profile. To examine it we will compare the H$\alpha$ flux with its
quiescent value.

The intensity of the Balmer emission lines of Z~And varies with the
orbital phase \citep{MK96,TTB08} and their flux during the
brightening must be compared with the quiescent one at the same
phase. We were not able to find in the literature Balmer fluxes at
orbital phase close to 0 in the quiescent state of Z~And.
\citet{MK96} have obtained the H$\alpha$ flux at phases 0.16 -- 0.34
where it increases from 40 $\times$ $10^{-12}$ to 103 $\times$
$10^{-12}$ erg\,cm$^{-2}$\,s$^{-1}$. \citet*{SSN97} have shown that
the Balmer emission lines of symbiotic stars in their quiescent
state are predominantly formed in the recombination zone which
separates the H$^{\rm o}$ from the H$^+$ region of the cool giant's
wind since the density of this zone is the highest one. These
authors have shown also that the Balmer lines are always
self-absorbed emission lines. These theoretical considerations show
that the absorption will be maximal at phase 0 where the giant is in
its inferior conjunction because of the highest density of the
absorbing particles. Then the line flux will have minimal value at
this phase and in the case of Z~And the H$\alpha$ flux will be
probably less than 40 $\times$ $10^{-12}$ erg\,cm$^{-2}$\,s$^{-1}$.
The comparison of this flux with fluxes (170 $\div$ 290) $\times$
$10^{-12}$ erg\,cm$^{-2}$\,s$^{-1}$ (Table~1) shows that it is
strongly increased during the brightening, which can be due to
addition of a new component in the line, possibly from an accretion
disc. The same conclusion is obtained for the line H$\gamma$ too.
The data of \citet{MK96} show that its flux at the orbital phase 0
is less than 2.1 $\times$ $10^{-12}$ erg\,cm$^{-2}$\,s$^{-1}$ and
the flux of the narrow component during the outburst is (16 $\div$
29) $\times$ $10^{-12}$ erg\,cm$^{-2}$\,s$^{-1}$ (Table~2).

Let us calculate the outer radius of the line emission region in the
disc from the shift between the two peaks. The shift was equal to
2.8 $\pm$ 0.2 \AA\, on November 12, 2002 at the time of the maximal
light. To calculate the outer radius the orbit inclination is
needed. There are several observational evidences supporting a high
orbit inclination of the system Z~And. Rayleigh scattering of the
hot radiation was detected in the spectrum of this system during its
1984 and 1985 outbursts \citep{FC95}. According to \citet{Sk03} the
far UV continuum is attenuated by Rayleigh scattering at orbital
phases close to 0 even in the quiescent state and an inclination of
76$^\circ$ was proposed on this base. The two-temperature type of
the UV spectrum observed during the 2000 -- 2002 outburst can be
interpreted supposing that the orbit inclination is rather high
\citep{Sk06}. On the other hand \citet{SS97} came to the conclusion
that the inclination is rather low proposing 47$^\circ$ $\pm$
12$^\circ$ obtained from their polarimetric orbit. Both points of
view are motivated and we cannot reject either of them. We suppose
also that the inclination can range from 47$^\circ$ to 76$^\circ$.
The results of our consideration do not depend strongly on this
quantity and we will perform our calculations with both of its
values keeping in mind that the actual inclination can be also
inside their range. At a mass of the hot component of 0.6
M$_{\odot}$ \citep{FC88,SS97} for the outer radius of the emission
region we obtain from 15 $\pm$ 2 R$_{\odot}$ to 26 $\pm$ 4
R$_{\odot}$ for orbit inclination of 47$^\circ$ -- 76$^\circ$. This
result supports our supposition, as it is in agreement with the
theoretical models of discs resulted from accretion of the stellar
wind \citep{Mitsumoto05}.

The intensity of the H$\alpha$ wings increased during the growth of
the light at the end of 2002. \citet{Skopal06} obtained synthetic
profiles arising from an optically thin bipolar stellar wind from
the hot components of the symbiotic stars and was able to make a fit
of the observed H$\alpha$ wings of a sample of ten symbiotic stars.
According to him, however, the profile formed in the stellar wind is
approximated by a function of the same type as that arising from
Raman scattering. For this reason he concluded that it is not
possible to distinguish between these two processes directly.
\citet{Ikeda04} noted that the polarization profile of the H$\alpha$
line of Z~And does not agree with that of its Raman $\lambda$ 6830
\AA\, line on October 25, 2002, i.e. during the brightening
considered by us. Based on the data of \citet{Ikeda04} we suppose
that the extended H$\alpha$ wings of Z~And during this brightening
are probably not due to Raman
scattered Ly$_\beta$ photons. Attention, however, should be paid to the fact that the FWZI of the H$\alpha$ wings of Z~And does not practically change during the active phase after the year 2000 and keeps its quiescent value of 4000 km\,s$^{-1}$. This fact gives us some reason to suppose that during the 2002 brightening the FWZI was determined mainly from radiation damping and the stellar wind can have some contribution in the wings at smaller distances from the centre of the line. The problem about the nature of these wings during active phase needs to be considered further.

        \subsubsection[]{The H$\gamma$ line}

The quiescent H$\gamma$ line of the system Z~And was of nebular type
and had an additional blue emission component with low intensity.
Its width was about 80--90 km\,s$^{-1}$ and occasionally exceeded
this value. The width remained the same during the 2000 -- 2002
brightening like the for line H$\alpha$ \citep{TTB08}.

During the growth of the light at the end of 2002 an additional
broad emission exceeding the local continuum by a factor of 1.3 and
with a full width at zero intensity (FWZI) of about 2000
km\,s$^{-1}$ appeared together with the nebular component of the
line (Figs.~2, 3). In this way the H$\gamma$ line consisted of a
narrow component of a nebular type and a broad component. At the
same time behavior of the narrow component closely followed that of
the H$\alpha$ line. Its profile was double-peaked and the width
increased by a factor of two compared with its quiescent value. The
dip between two peaks of the line had the same radial velocity at
all times of observations, which was close to the velocity of the
dip of the H$\alpha$ line (Fig.~4).

Let us calculate the outer radius of the emission region of the line
in the disc using the shift between the two peaks. This shift was
equal to 1.4 $\pm$ 0.1 \AA\, on November 12 and with use of the
parameters of the system adopted by us it gives an outer radius from
26 $\pm$ 3 R$_{\odot}$ to 46 $\pm$ 6 R$_{\odot}$ for orbit
inclination of 47$^\circ$ -- 76$^\circ$. This radius is in agreement
with the theory of discs formed as a result of wind accretion like
the size of the H$\alpha$ region \citep{Mitsumoto05}. We obtained
that the outer radius of the emission region of the H$\gamma$ line
is greater than that of H$\alpha$ but when atoms are excited as a
result of recombination and the density decreases with the distance
from the star, the region of the H$\alpha$ line is expected to be
located outward from that of H$\gamma$. Our result follows from the
greater separation of the peaks of the line H$\alpha$.

The data propose not only greater peaks' separation, but also greater width of this line. If the emitting region is fully transparent in Balmer lines their width will be the same. The Balmer lines of symbiotic stars, however, are always self-absorbed emission lines. The reason for the greater width and peaks' separation of H$\alpha$ of Z~And can be its higher optical depth compared to H$\gamma$. Then for the size of the line emission region in the disc we will prefer mostly the results based on the H$\gamma$ data.

There is, however, an other possibility for the greater separation of the peaks of the line H$\alpha$ more, which follows from the flow structure. This line is possible to
be emitted at greater distance from the orbital plane, but more
close to the axis of the disc, where the rotational velocity is
greater. In this way it can be strongly contaminated by emission of
the disc-like envelope surrounding the accretion disc. This envelope
forms from the material ejected during the previous outburst and 
fallen back after its finish. The conservation of the angular
momentum of the ejected material is responsible for its formation.
Close to the axis of rotation the velocity is large enough to
explain the greater width of the line H$\alpha$ and the greater
separation of its peaks compared to the line H$\gamma$.

\begin{figure*}
    \resizebox{18pc}{!}{\includegraphics[]{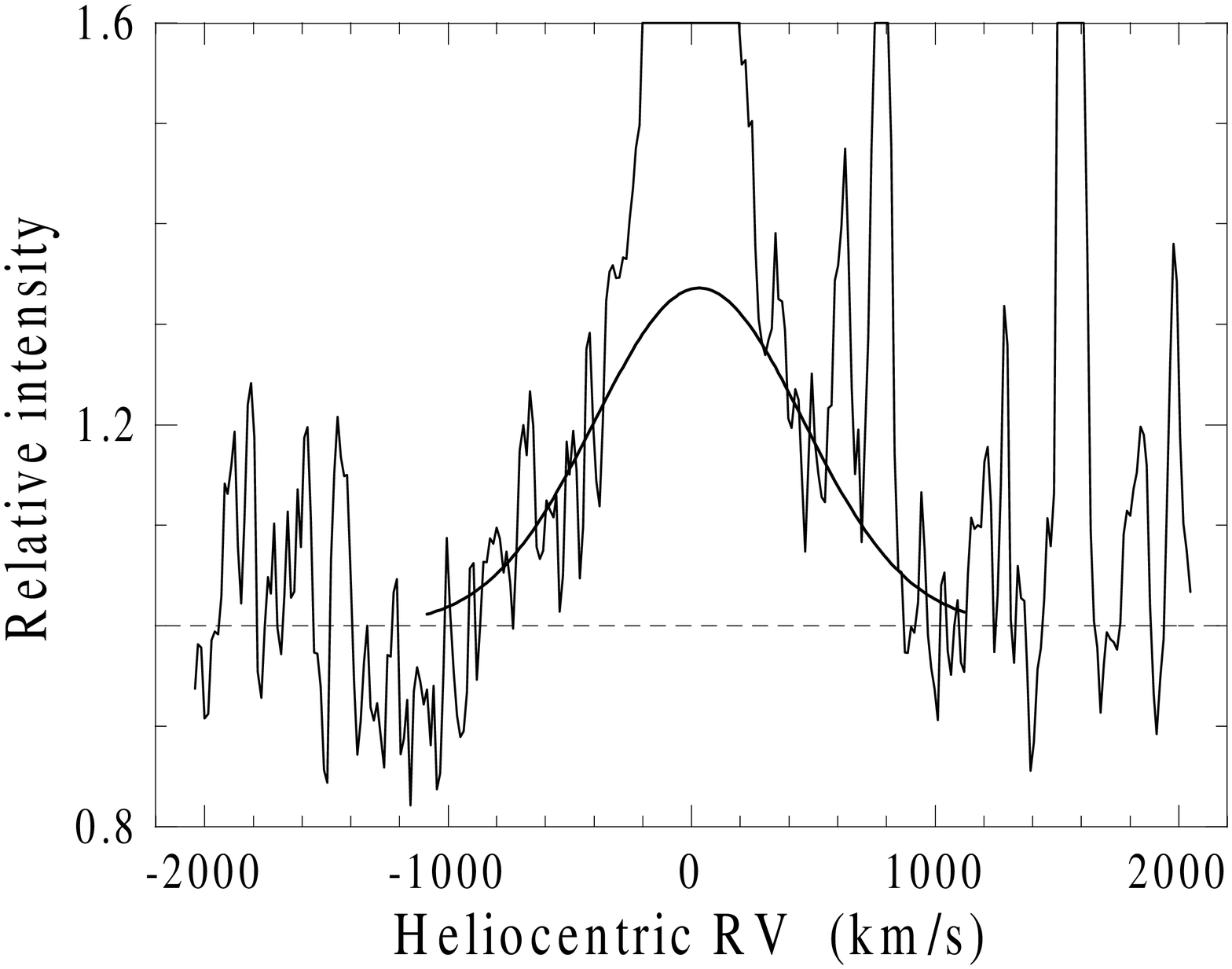}}
    \hspace{0.5cm}
    \resizebox{18pc}{!}{\includegraphics[]{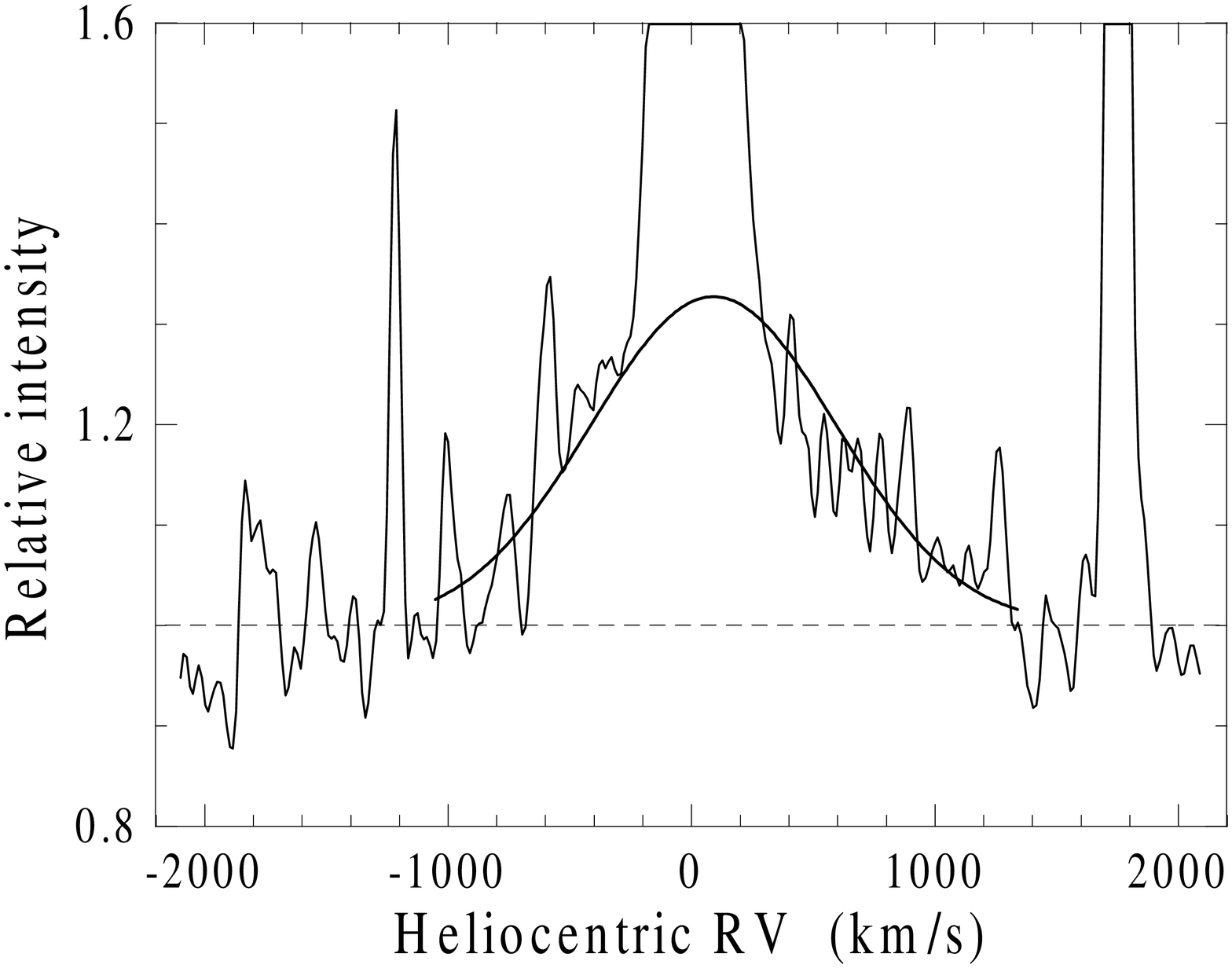}}

{\bf Figure 3.} The H$\gamma$ (left panel) and \mbox{He\,{\sc ii}} $\lambda$\,4686 (right panel) broad components.
The level of the local continuum is marked with a dashed line.

\label{HgHe2_fit}
\end{figure*}

The broad component of the H$\gamma$ line was measured in the
following way. The observed spectrum was corrected through removing
several weak emission lines of \mbox{Fe\,{\sc ii}}, \mbox{O\,{\sc
ii}} and \mbox{N\,{\sc iii}} as well as the most intensive
absorption lines of the giant. After that it was analyzed by fitting
with a Gaussian function (Fig.~3) and its parameters obtained with
this procedure are listed in Table~2. The error of the equivalent
width is not greater than 20 per cent depending on the noise of the
different spectra.

\begin{table*}

{\bf Table 2.} The data of the lines H$\gamma$ and \mbox{He\,{\sc ii}} $\lambda$\,4686. N and B denote narrow and
broad component respectively. The flux is in units $10^{-12}$ erg\,cm$^{-2}$\,s$^{-1}$ and the other quantities are
in units km\,s$^{-1}$.

\label{HgammaHe2}

 \begin{tabular}{@{}|c|cc|cc|cc|cc|c|cc|}
  \hline
    Date & \multicolumn{2}{c}{FWHM(N)} & \multicolumn{2}{c}{Flux(N)} & \multicolumn{2}{c}{FWHM(B)} & \multicolumn{2}{c}{FWZI(B)} & \multicolumn{1}{c}{$\upsilon_{\rm {h}}$} & \multicolumn{2}{c|}{Flux(B)} \\
    \cline{2-12}
             & H$\gamma$ & \mbox{He\,{\sc ii}} & H$\gamma$ & \mbox{He\,{\sc ii}} & H$\gamma$ & \mbox{He\,{\sc ii}} & H$\gamma$ & \mbox{He\,{\sc ii}} & \mbox{He\,{\sc ii}} & H$\gamma$ & \mbox{He\,{\sc ii}} \\

    \hline
 Sept. 25 & 166 & ~87 & 15.718 & 46.319 & 1014 $\pm$ 100 & 1091 $\pm$ ~~92 & 1900 & 2113 & 1050 & 3.193 & 3.713 \\
 Oct. 20~ & 165 & ~92 & 22.796 & 59.782 & 1073 $\pm$ ~~78 & 1170 $\pm$ 101 & 2100 & 2201 & 1100 & 4.885 & 5.234 \\
 Nov. 12 & 171 & 105 & 28.800 & 70.110 & 1010 $\pm$ ~~70 & 1192 $\pm$ ~~56 & 2030 & 2416 & 1200 & 5.236 &6.658 \\
      \hline
 \end{tabular}

\end{table*}

To conclude about the nature of the gaseous environment where the
broad component appeared we will treat different mechanisms of line
broadening. One of them is the electron scattering. The total flux
of the line, which is a sum of the fluxes of two components, is (19
-- 34) $\times$ 10$^{-12}$ erg\,cm$^{-2}$\,s$^{-1}$ (Table~2). This
gives emission measures of (1 -- 2) $\times$ 10$^{59}$ (d/1.12
kpc)$^{2}$ cm$^{-3}$. To calculate the radius of a spherical
emitting volume, having this emission measure, the mean electron
density is needed. \citet{FC88} obtained a mean quiescent electron
density of 10$^{10}$ cm$^{-3}$ in the nebula of Z~And. We assume
that the mean electron density in the close vicinity of the compact
object during the brightening is close to the quiescent electron
density and will perform our calculations with the value 10$^{10}$
cm$^{-3}$. We come to the same conclusion, however, if we use lower
densities too. We obtain a radius of (6 -- 8) $\times$ 10$^{12}$
(d/1.12 kpc)$^{2/3}$ cm of the spherical emitting volume. If the
broad component appeared due only to the electron scattering it
would be risen in region with optical thickness of 0.17 -- 0.19.
Using these values and a density of 10$^{10}$ cm$^{-3}$, we derive
the radius of almost 3 $\times$ 10$^{13}$ (d/1.12 kpc)$^{2/3}$ cm,
corresponding to an enormous emission measure of 10$^{61}$ (d/1.12
kpc)$^{2}$ cm$^{-3}$. This result differs from the previous one and
we conclude that the broad component is probably not produced by
electron scattering.

\begin{figure*}
    \includegraphics[height=.32\textheight]{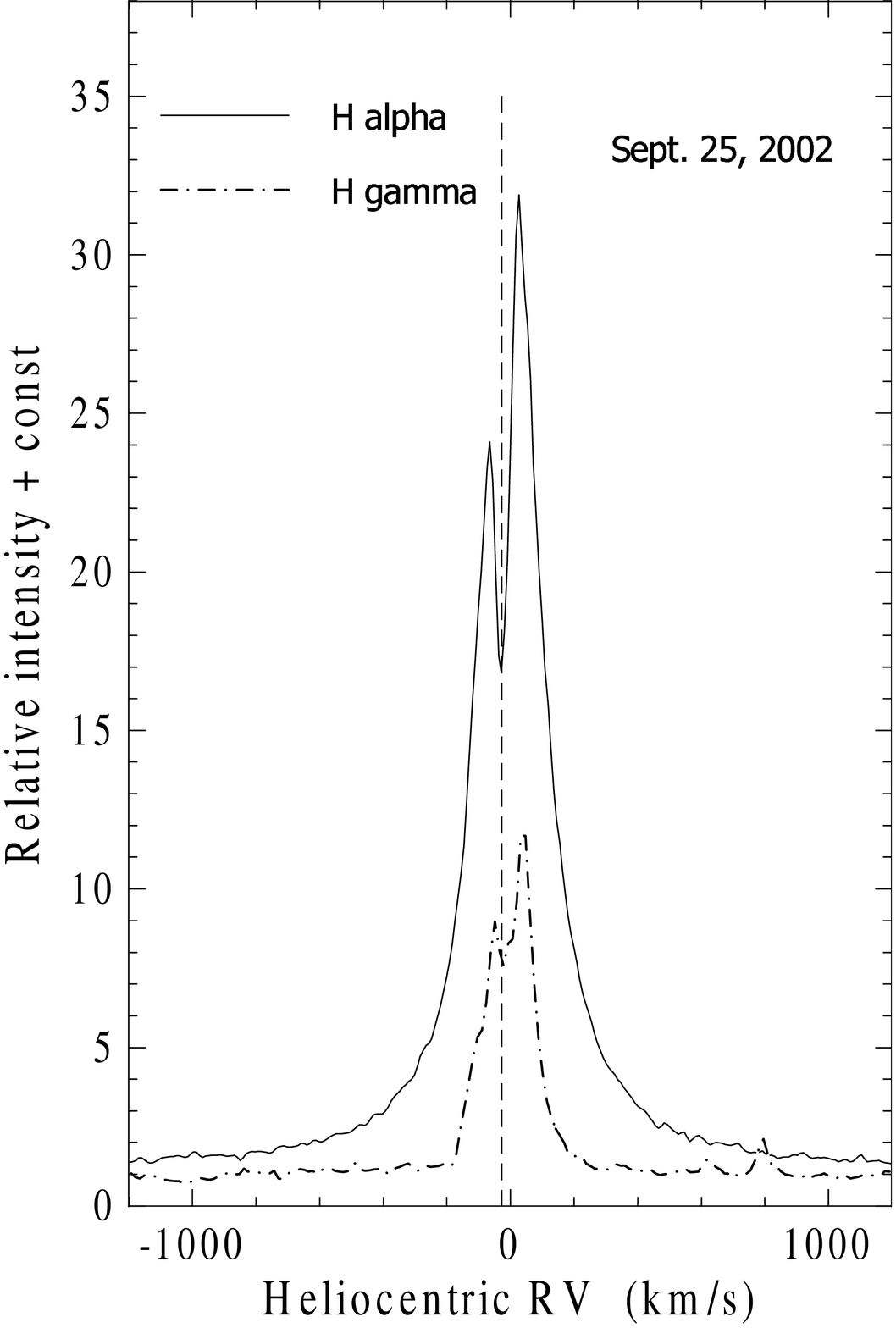}
    \includegraphics[height=.32\textheight]{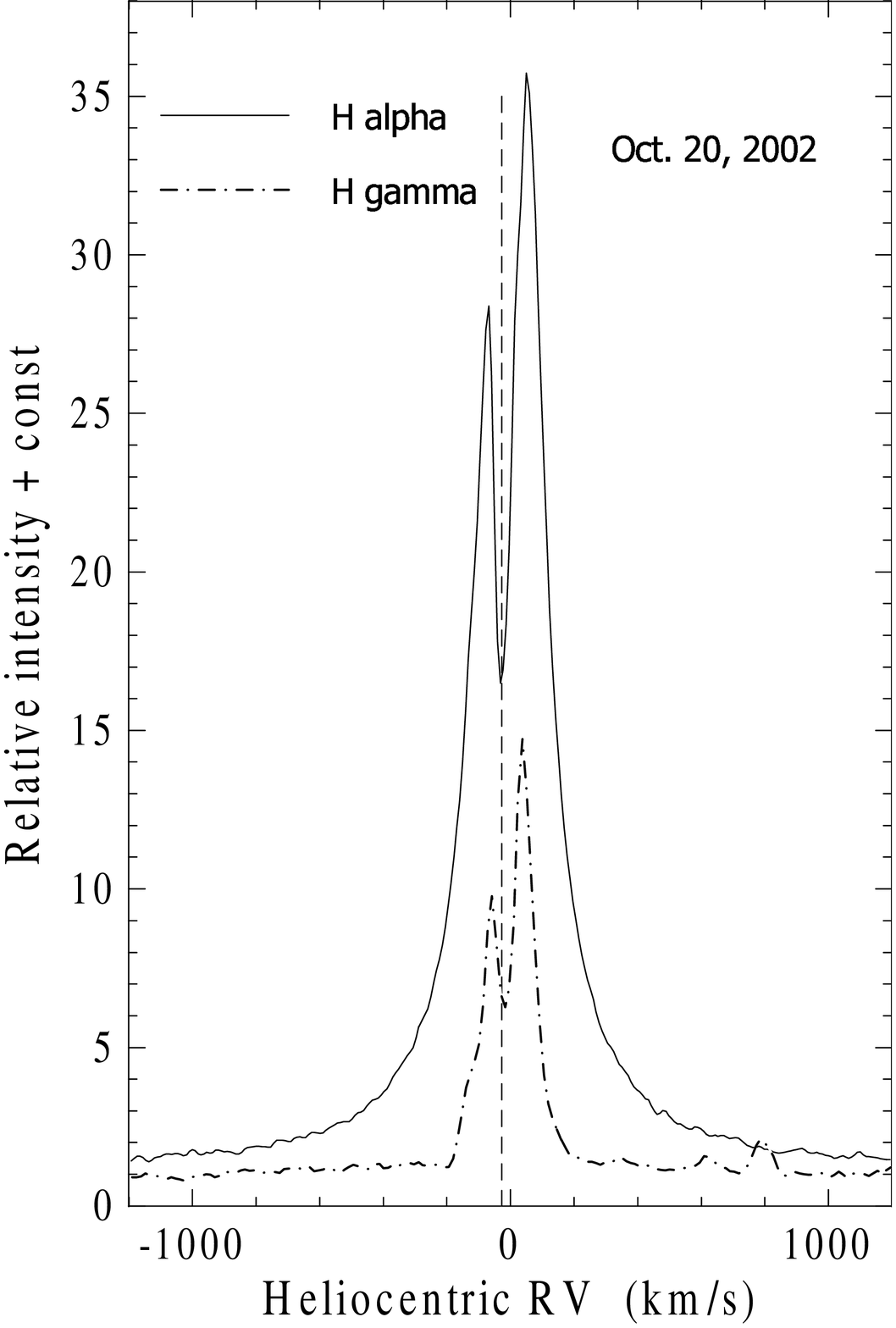}
    \includegraphics[height=.32\textheight]{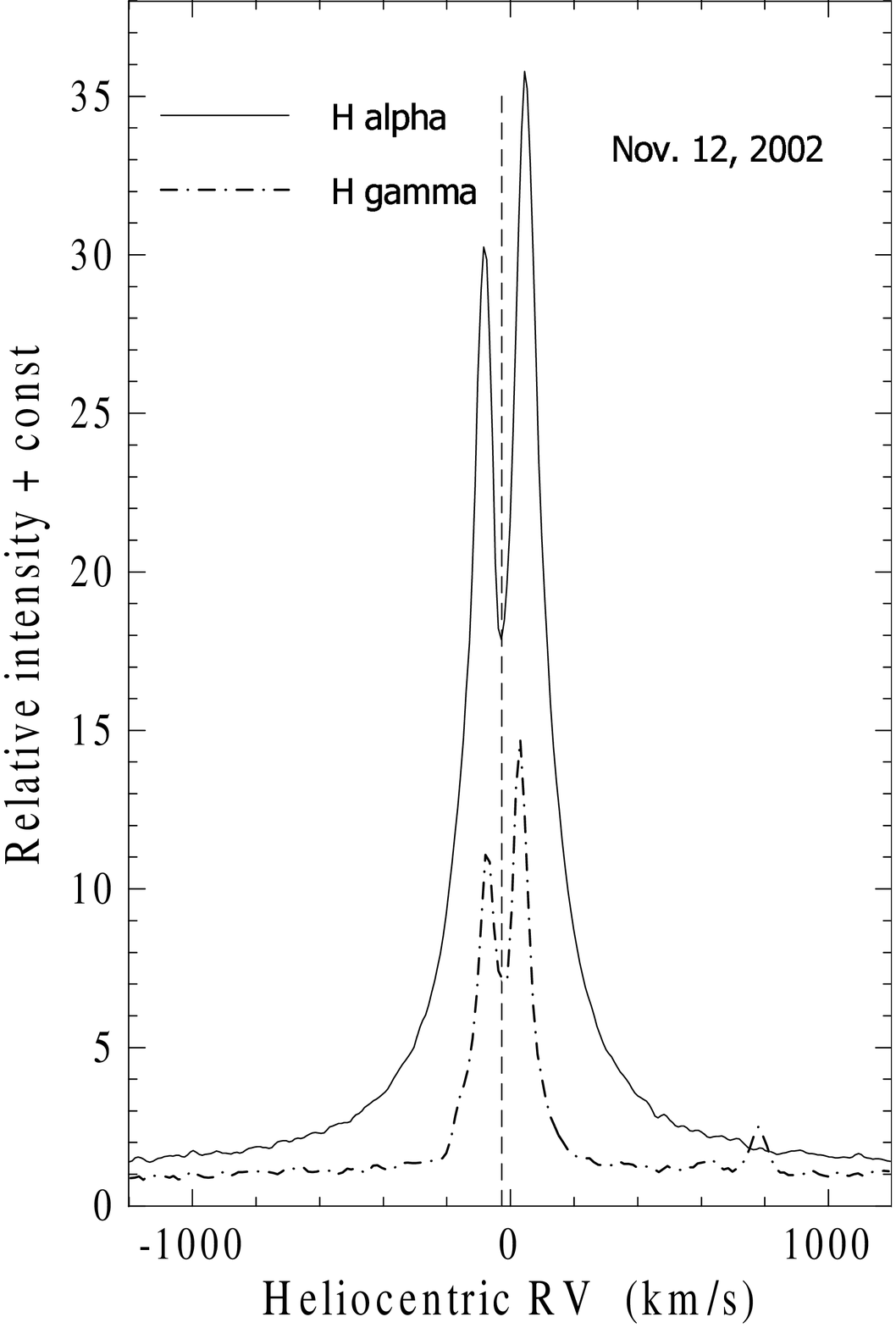}

{\bf Figure 4.} The lines H$\alpha$ and H$\gamma$ in each time of observation. The dips of the two lines have close
velocity positions.

\label{Ha&Hg}
\end{figure*}

An other possible mechanism the broad component to appear is to be emitted by an
optically thin Keplerian accretion disc. The half width at zero
intensity of the broad component on November 12 at the time of the
light maximum was about 1020 km\,s$^{-1}$ (Table~2). The velocity, derived
from the width at zero intensity of the line, is related to the
movement at the inner boundary of the disc. This velocity, however,
is on the line of sight. Taking into consideration the orbit
inclination of 47$^\circ$ -- 76$^\circ$, the obtained velocity at
the inner boundary is 1400 -- 1050 km\,s$^{-1}$. With the mass of
the compact object adopted by us $M_{wd}=0.6M_{\odot}$ we obtain an
inner radius of the disc of $0.06 \div 0.10$ R$_{\odot}$. This estimate is
smaller than the upper limit of the radius of this
object at the time of the maximal light ($\sim$ 0.13\,(d/1.12kpc)
R$_{\odot}$ evaluated by \citet{TTT04}).
The estimate of the disc radius, however, has an appreciable uncertainty due to the velocity, which depends from the level of the local continuum. Our data take small range of 200 \AA\, which leads to additional difficulty in determining continuum level. Its uncertainty can reach up to 5 per cent. Variation of 5 per cent can lead to reduction of 20 per cent of the velocity which is based on the FWZI of the line. So, a velocity of 800 km\,s$^{-1}$ gives us an inner radius of the disc of 0.09 $\div$ 0.17 R$_{\odot}$. Then we can think that the inner radius is about 0.1 $\div$ 0.2 R$_{\odot}$ for the range of orbit inclination treated by us and is in satisfactory agreement with the upper limit of the radius of the compact object of 0.13(d/1.12 kpc)R$_{\odot}$. That is why we will suppose that the emission of an optically thin accretion disc can be possible reason for the broad H$\gamma$ component.

        \subsection[]{The \mbox{He\,{\sc ii}} $\lambda$\,4686 line}

During the rise of the light the \mbox{He\,{\sc ii}} $\lambda$\,4686
line  (Fig.~5) consisted of two emission components -- a narrow central
component with a width of about 90 -- 100 km\,s$^{-1}$ and a broad
component whose width was much greater (Fig.~3). The broad component
was analyzed in the following way.

\begin{figure}

   \includegraphics[height=.45\textheight]{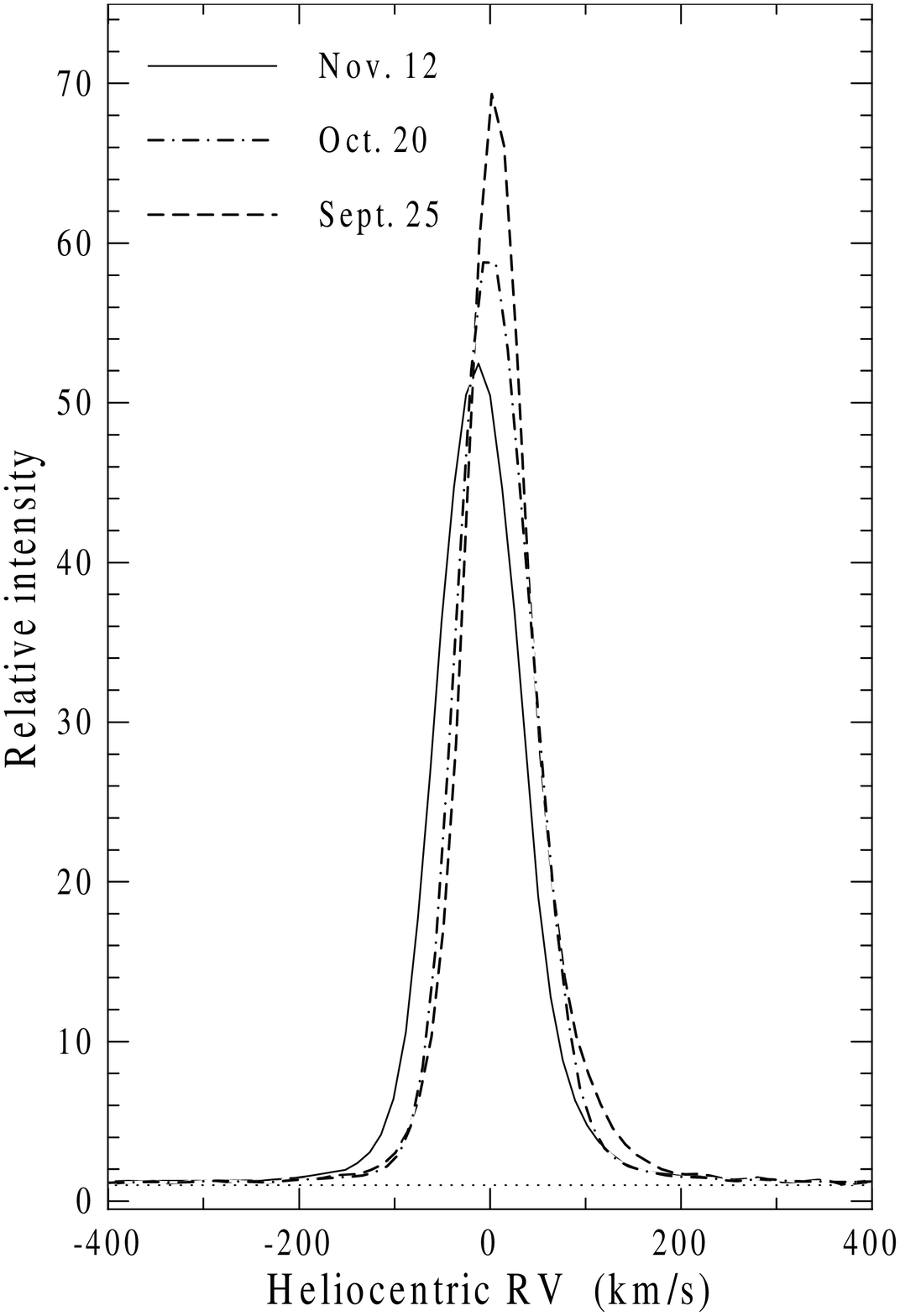}

{\bf Figure 5.} The profile of the \mbox{He\,{\sc ii}} $\lambda$\,4686 line. The level of the
local continuum is marked with a doted line.

\end{figure}

The observed spectrum was corrected with removal of several weak
emission lines of \mbox{O\,{\sc ii}} and the strongest absorption
lines of the giant. Then it was fitted with a Gaussian function. The
parameters obtained with this procedure are presented in Table~2.
The error of the equivalent width ranges from 11 to 19 per cent
depending on the noise of the individual spectrum.

The narrow central component of the line \mbox{He\,{\sc ii}}
$\lambda$\,4686 was probably emitted in the ionized portion of the
giant's wind and in that part of the disc's corona where the high
velocity wind does not propagate as well.

As in the case of the line H$\gamma$ we must examine the possibility
the broad component to be formed by electron
scattering. The total flux of the line is (50 -- 77) $\times$
10$^{-12}$ erg\,cm$^{-2}$\,s$^{-1}$ (Table~2) which leads to
emission measures of about 2 $\times$ 10$^{59}$ (d/1.12 kpc)$^{2}$
cm$^{-3}$. Using the mean density of 10$^{10}$ cm$^{-3}$
\citep{FC88} we obtain the radius of (6 -- 8) $\times$ 10$^{12}$
(d/1.12 kpc)$^{2/3}$ cm of the spherical emitting volume. If the
broad component was determined only by electron scattering the
optical thickness of its emission region would be 0.08 -- 0.09.
These values and the density of 10$^{10}$ cm$^{-3}$ give the radius
of more than 10$^{13}$ (d/1.12 kpc)$^{2/3}$ cm, corresponding to
great emission measures of (6 -- 10) $\times$ 10$^{59}$ (d/1.12
kpc)$^{2}$ cm$^{-3}$. This result differs from the previous one and
we conclude on this base that the broad component is probably not
determined by the electron scattering.

The red wing of the \mbox{He\,{\sc ii}} $\lambda$\,4686 broad component at all spectra is more extended than the blue one causing one general asymmetry in this line, which however, is absent in the H$\gamma$ broad component (see Fig.~3). This asymmetry give us a reason to suppose that the line \mbox{He\,{\sc ii}} $\lambda$\,4686 does not appear in an accretion disc but rather -- in a region of a high velocity stellar wind. It is seen in Table~2 that the FWZI of the H$\gamma$ broad component is always smaller than that of the \mbox{He\,{\sc ii}} $\lambda$\,4686 broad component. These data propose that the line H$\gamma$ is not emitted in the wind region of the \mbox{He\,{\sc ii}} $\lambda$\,4686 line. If the line H$\gamma$ was emitted in more outer region of the wind, its FWZI would not be smaller than that of the \mbox{He\,{\sc ii}} $\lambda$\,4686 line, since the velocity in this region is not smaller than in the \mbox{He\,{\sc ii}} $\lambda$\,4686 region. Consequently the line H$\gamma$ is not emitted in the wind. Then these two lines are emitted probably in different regions of the binary system -- H$\gamma$ in the accretion disc and \mbox{He\,{\sc ii}} $\lambda$\,4686 in the stellar wind.

    \section[]{Mass-loss rate}

The mass-loss rate of the hot compact companion was calculated with use of the energy flux of the broad component of the line \mbox{He\,{\sc ii}} $\lambda$\,4686 in the same way as in our previous work \citet{TTB08}. It was supposed that the outflow has a spherical symmetry and a constant velocity. In our calculations we used a wind velocity based on the half width at zero intensity of the line from Table~2. This velocity is actually an arithmetical mean of the velocities of the two wings. It was also supposed that the wind is fully transparent in this line. We adopted an electron temperature of 30\,000 K in the wind and a distance to the system d $=1.12$ kpc \citep{FC88,FC95} as in the case of the major brightening. The line is supposed to be emitted by a spherical layer and the radii of integration are need. The inner radius is the stellar radius. We calculated the mass-loss rate for the time November 12 at the light maximum since we had data for the stellar radius only for that time. We used the upper limit of the radius of 0.13\,(d/1.12kpc) R$_{\odot}$ from the work of \citet{TTT04} and so we can obtain the upper limit of the mass-loss rate. The outer radius of integration is equal to infinity. A recombination coefficient for case B \citep{SH} corresponding to temperature of 30\,000 K and the density  at the level of the photosphere was used. An upper limit of the rate of 1.8 $\times$ 10$^{-7}$ (d/1.12 kpc)$^{3/2}$ M$_{\odot}$\,yr$^{-1}$ was derived with an uncertainty of about 30 per cent due to the observational data. The real uncertainty, however, is determined from use of a model with a constant velocity of the wind too and can be higher. We obtained in this way that the mass-loss rate at the time of the light maximum is close to that of 2.4 $\times$ 10$^{-7}$ (d/1.12 kpc)$^{3/2}$ M$_{\odot}$\,yr$^{-1}$ at the time of the light maximum of the major 2000 -- 2002 brightening.

    \section[]{Discussion}

An optically thin accretion disc which can give rise to the
double-peaked Balmer profiles, could appear at the final stage of
the major 2000 -- 2002 brightening. At that time the residual
accretion disc still existed in the system. In addition some part of
the ejected mass was in the potential well of the compact object.
After the discontinuation of the flow the accretion of this material
began again which led to a strong increase of the accretion rate.
That part of the ejected mass which has had orbital angular momentum
great enough, formed the disc itself and a disk-like envelope. An
accretion disc of such a type is possible to exist in the system
during all brightenings following the major ones.

The growth of the optical light of the system during the
small-amplitude brightening was mainly due to the nebular emission
\citep{TTT04}. If the accretion disc is mainly responsible for the
Balmer lines its contribution in the nebular continuum of the system
will be the greatest one and its emission will determine the optical
maximum. The disc emission increases not only as a result of an
additional input of mass but probably because of the growth of the
Lyman luminosity of the ionizing source too. The supposition for a
growth of this luminosity is supported by the following
observational fact. The emission measure of the nebula increased by
a factor of 2.5 $\pm$ 0.2 comparing to the quiescent state of the
system \citep{TTT04} which means that the number of recombinating
ions has increased in the same ratio. This is probably due not only
to the collisional ionization but to the radiative one too. The
increase of the radiative ionization is determined by change of the
Lyman luminosity of the ionizing star.

The double-peaked profiles of the hydrogen lines give a reason to
suppose that the lines can be emitted mainly by an optically thin
accretion disc. Using the H$\gamma$ line we estimated the outer
radius of the accretion disc as 26 $\pm$ 3 $\div$ 46 $\pm$ 6 R$_{\odot}$
for the orbit inclination angle in a range from 47$^\circ$ to
76$^\circ$. We tryed to obtain one approximate estimate of 
the inner radius of the disc too, which amounts to about 0.1 $\div$ 0.2 R$_{\odot}$ 
and does not contradict to the supposition of
\citet{SB99} for magnetosphere of the white dwarf in Z~And.

Our supposition for the presence of an accretion disc and a high
velocity stellar wind in the system gives a reason to assume that
its visual line emission spectrum can be explained in the framework
of the model for the interpretation of the spectral behavior during
the major 2000 -- 2002 brightening \citep{TTB08}. This model
provides an explanation of the observed two-velocity regime of the
mass outflow when a P~Cyg wind with a low-velocity was observed
together with the high-velocity optically thin wind. The
low-velocity of the P~Cyg absorption component was interpreted with
slowing down the wind because of its collision with the disc. A
P~Cyg absorption component was absent in the spectrum during the
brightening -- subject of research in this paper. To observe
component of such a type the column density of the absorbing gas
needs to be high enough and the gas to be projected on the observed
photosphere as well. During the small-amplitude brightening the
column density in the wind was close to that during the major
brightening. Then the possible reason can be that the gas,
outflowing after the collision is not projected on the observed
photosphere.

    \section[]{Conclusion}

We present results of high-resolution observations of the Balmer
lines H$\alpha$ and H$\gamma$ as well as the line \mbox{He\,{\sc
ii}} $\lambda$\,4686 of the symbiotic binary Z~And during the rise
of the light of its small-amplitude brightening at the end of 2002.
The width of the Balmer lines was greater by a factor of two than in
the quiescent state of the system and their profiles were
double-peaked.

We compared the emission measure of the circum-binary nebula of the
system and fluxes of the two Balmer lines at the times of the light
maxima in December 2000 and November 2002 which were at close
orbital phases. We assumed close physical conditions in the nebula
at those times. It was concluded on this base that the optical depth
in the lines was greater in December 2000. This conclusion proposes
that the double-peaked profile during the 2002 brightening can be
explained not only with the self-absorption, but mainly with the
velocity distribution of the emitting particles. The most probable
reason can be the rotation of the accretion disc with the outer
radius lying in a range 26 $\pm$ 3 $\div$ 46 $\pm$ 6 R$_{\odot}$ for orbit
inclination angle from 47$^\circ$ to 76$^\circ$.

The H$\gamma$ line had a broad emission component showing a velocity of about 1000 km\,s$^{-1}$. Assuming that it is emitted in the inner part of the accretion disc we estimated rougly the inner radius which amounts to 0.1 $\div$ 0.2 R$_{\odot}$ for the same orbit inclination. 

The line \mbox{He\,{\sc ii}} $\lambda$\,4686 had a broad emission component whose FWZI was always greater than that of the line H$\gamma$. This component was assumed to be due to high velocity stellar wind from the hot compact component of the system.

We suppose that an optically thin accretion disc, giving rise to
double-peaked profile existed in the system during the brightening
observed by us (and possibly many other brightenings following the
major ones) since there were good conditions for its formation. At
the final stage of the major 2000--2002 brightening some part of the
ejected mass was in the potential well of the compact object. After
the discontinuation of the flow the accretion of this material began
again. This part of the ejected mass which had orbital angular
momentum great enough, formed the disc.

The H$\alpha$ line had broad wings extended to not less than 2000 km\,s$^{-1}$ from its center.
We supposed that they were determined mainly from radiation damping but stellar wind could contribute to their emission at smaller distance from the centre of the line.

Assuming that the broad component of the line \mbox{He\,{\sc ii}} $\lambda$\,4686 
is due to a high velocity stellar wind we estimate the
upper limit of the mass-loss rate of the companion resulting
from this wind as
$\sim$1.8 $\times$ 10$^{-7}$ (d/1.12 kpc)$^{3/2}$ M$_{\odot}$\,yr$^{-1}$.

Based on this interpretation we conclude that the behavior of the
spectral lines can be explained in the framework of a model of the
system proposed for its major 2000 -- 2002 brightening where the
high velocity stellar wind of the compact object collides with the
accretion disc \citep{TTB08}.

\section{Acknowledgements}

The authors are thankful to the referee Acad.~A.~Boyarchuk for his helpful comments and discussion which contributed to improve the paper.
The work was partly supported by funds under
agreement between Bulgarian Academy of Sciences and Russian Academy of
Sciences on Fundamental Space Research (Project 2.11).
NAT's contribution to this work was supported in part by Bulgarian National Science Foundation grant under contract DO 02-85.
DVB was supported by the
Russian Academy of Science, the Russian Foundation for
Basic Research (projs. 08-02-00371, 09-02-00064), by
the Federal Agency of Science and Innovations, and by The Federal Targeted Programme 
"Scientific and Educational Human Resources of Innovation-Driven Russia" for 2009-2013.

\end{document}